# Investigating The Impacting Factors on The Public's Attitudes Towards Autonomous Vehicles Using Sentiment Analysis from Social Media Data

**Shengzhao Wang**, Master Student
Key Laboratory of Road and Traffic Engineering of the Ministry of Education
College of Transportation Engineering, Tongji University
4800 Cao'an Highway, Shanghai, 201804, China
E-mail: wszwsz179@163.com

**Meitang Li**, Research Assitant
University of Michigan Transportation Research Institute
2901 Baxter Rd, Ann Arbor, MI, USA, 48109-2150
E-mail: meitang@umich.edu

**Bo Yu**, Ph.D., Assistant Professor, Corresponding Author*
Key Laboratory of Road and Traffic Engineering of the Ministry of Education
College of Transportation Engineering, Tongji University
4800 Cao'an Highway, Shanghai, 201804, China
E-mail: boyu@tongji.edu.cn

**Shan Bao,** Ph.D., Associate Professor
Industrial and Manufacturing Systems Engineering Department, University of Michigan-Dearborn, 4901 Evergreen Rd, Dearborn, MI 48128
University of Michigan Transportation Research Institute
2901 Baxter Rd, Ann Arbor, MI, USA, 48109-2150
E-mail: shanbao@umich.edu

**Yuren Chen**, Ph.D., Professor
Key Laboratory of Road and Traffic Engineering of the Ministry of Education
College of Transportation Engineering, Tongji University
4800 Cao'an Highway, Shanghai, 201804, China
E-mail: chenyr@tongji.edu.cn

**Abstract**

The public's attitudes play a critical role in the acceptance, purchase, use, and research and development of autonomous vehicles (AVs). To date, the public's attitudes towards AVs were mostly estimated through traditional survey data with high labor costs and a low quantity of samples, which also might be one of the reasons why the influencing factors on the public's attitudes of AVs have not been studied from multiple aspects in a comprehensive way yet. To address the issue, this study aims to propose a method by using large-scale social media data to investigate key factors that affect the public's attitudes and acceptance of AVs. A total of 954,151 Twitter data related to AVs and 53 candidate independent variables from seven categories were extracted using the web scraping method. Then, sentiment analysis was used to measure the public attitudes towards AVs by calculating sentiment scores. Random forests algorithm was employed to preliminarily select candidate independent variables according to their importance, while a linear mixed model was performed to explore the impacting factors considering the unobserved heterogeneities caused by the subjectivity level of tweets. The results showed that the overall attitude of the public on AVs was slightly optimistic. Factors like "drunk", "blind spot", and "mobility" had the largest impacts on public attitudes. In addition, people were more likely to express positive feelings when talking about words such as "lidar" and "Tesla" that relate to high technologies. Conversely, factors such as "COVID-19", "pedestrian", "sleepy", and "highway" were found to have significantly negative effects on the public's attitudes. The findings of this study are beneficial for the development of AV technologies, the guidelines for AV-related policy formulation, and the public's understanding and acceptance of AVs.

# 1. Introduction

The advent of autonomous vehicles (AVs) represents a new technological revolution in the transportation sphere. AVs have a great potential in reducing gas emissions, fuel consumption, and traffic accidents associated with human-related errors, and increasing mobility especially for the elderly and the disabled (Payre et al., 2014; Combs et al., 2019; Bennett et al., 2019). However, the popularity of AVs is still at a low level due to the legal, liability, privacy, and safety issues (Fagnant and Kockelman, 2015). It was shown in a survey study that 63% of the respondents were not likely to use AVs (Abraham et al., 2017). The public's attitudes play a critical role in the acceptance, purchase, use, and research and development of AVs (Menon et al., 2016; Das et al., 2019; Nordhoff et al., 2018).

Although many studies have tried to explore factors affecting peoples' acceptance of AVs, most of them used traditional survey methods, such as online questionnaires (Pettigrew et al., 2019; Bansal et al., 2016) and field investigation (Kassens-Noor et al., 2020; Bennett et al., 2019). For example, a national online survey with open-ended questions was administered to a sample of 1,624 Australians, and it showed that the negative attitudes towards AVs may be derived from a combination of cognitive and emotional factors (Pettigrew et al., 2019). However, the sample size of the traditional survey method is relatively small in the range of 300-5,000 copies, which may be not able to effectively represent the public's attitudes (Gkartzonikas and Gkritza, 2019). Additionally, labor cost consumption and limitations of time and space are also additional weaknesses of the questionnaire survey regarding transportation services (Liu et al., 2019).

These survey studies on the impacting factors of the attitudes towards AVs found that: (1) Males, younger people, urban residents, and those with higher income and higher levels of education were likely to hold positive attitudes towards AVs (Nielsen and Haustein, 2018; Hulse et al., 2018; Hardman et al., 2019). (2) Better fuel and energy efficiency and lower vehicle emissions were expected to be brought from AVs by those who pursued environmental protection (Begg, 2014; Casley et al., 2013). (3) The unwillingness to pay any additional fees for AVs was expressed by a large proportion of respondents (Daziano et al., 2017). (4) Safety of AVs, security issues, traffic congestion, and automatic parking were also reported to be impacting factors of public acceptance towards AVs (Schoettle and Sivak, 2014; Payre et al., 2014).

Unlike traditional survey methods, social media provides a new source of timely data at a large scale and low cost. As of the first quarter of 2021, Twitter (a representative platform of social media) had 199 million monetizable daily active users worldwide. It was entirely possible to obtain wide-range information from the public promptly by social media (Zhang et al., 2018). People had become both producers and disseminators of news and reviews with the flourishment of social media (Rajendran and Thesinghraja, 2014). It can be found that a large number of comments related to AVs have been released on social media, and the public's close interest is essential for the promotion and application of AV technologies.

There have been some studies using social media data to perform further analysis in the field of transportation. For example, the 1-year over 3 million Twitter contents in Northern Virginia and New York City were extracted to detect traffic accidents (Zhang et al., 2018). The results showed that nearly 66% of the accident-related tweets can be located by the actual accident on highways, indicating the possibility to regard tweets as an accident detection tool due to their effectiveness of time and location. In addition, due to the large scale and real-time of social media data, these kinds of new data had also been introduced in the studies of traffic flow prediction (Lin et al., 2015; Ni et al., 2016), traffic strategy designation (Cottrill et al., 2017), route planning (Huang et al., 2017), etc.

To analyze the big data from social media, natural language processing (NLP) is a potentially powerful method. NLP is an important direction in the field of computer science, which provides an opportunity to understand people's opinions regarding hot topics by enabling computers to obtain the meaning of human language from text documents (Anta et al., 2013; Jelodar et al., 2021). In the field of transportation, various objectives such as traffic investigation, safety, and management can be achieved in the NLP methods (Stambaugh, 2013). Sentiment analysis was one of the most commonly-used methods in NLP for analyzing attitudes and feelings (Liu, 2012). In a study by Liu et al. (2019), a sentiment analysis method was proposed to calculate people's satisfaction with different transit facilities based on the data extracted from the website. In another study, the sentiment analysis was performed for analyzing the comments on the 15 most-viewed AV-related videos (Das et al., 2019).

Many different statistical analyses and machine learning methods have been used to investigate the impacting factors on public attitudes towards AVs, such as t-tests and ANOVA (Pettigrew et al., 2019), support vector machines algorithm (Kohl, et al., 2017), the univariate model (Bansal et al., 2016), and so on. Compared to the above methods, the mixed model can consider unobserved heterogeneities to obtain a more accurate relationship of variables (Wang et al., 2017a). It had been employed in many studies to explore significant factors. For example, mixed model analyses were used to assess and test the impacting factors on drivers' compliance with the speed choice suggestions (Yu et al., 2019a). Driver behavior in response to the warning was modeled by mixed-effects Poisson regression models (Jermakian et al., 2017).

Additionally, since social media data can provide abundant variables, variable selection is needed at the beginning of analyses. The Random forests (RF) algorithm is widely employed to screen variables in the early stage because it can provide the variables' importance and handle a large number of variables with higher dimensionality at the same time (Wright and Ziegler, 2015). For instance, the most important explanatory variables associated with severe crash occurrence were selected by RF (Yu and Abdel-Aty, 2014). RF was also used for the selection of significant variables related to the injury severity of secondary incidents (Li et al., 2020a).

Given the above, most current studies have explored the public's attitudes towards AVs through questionnaires. However, the quantity of data collected in the traditional survey methods was very limited, and the influencing factors on the public's attitudes and acceptance of AVs have not been studied from multiple aspects in a comprehensive way yet. To fill this research gap, this study aims to use larger and more timely social media data to investigate the public's attitudes towards AVs and their impacting factors.

In this study, a total of 954,151 tweets and 53 candidate independent variables were extracted using the web scraping method, and the sentiment analysis method was used to measure public attitudes on AVs. Then, the Random forest algorithm and a linear mixed model were employed to analyze the impacting factors. The results of this study are beneficial for the development of AV technologies, the guidelines for AV-related policy formulation, and the public's understanding and acceptance of AVs.

## 2. Methodology
### 2.1 Data extraction

In this study, a sentiment analysis method was applied to investigate the public's attitudes towards AVs and their impacting factors based on large-scale Twitter data. Sentiment analysis is a ramification in the natural language processing (NLP) domain, which focuses on detecting, processing, and analyzing the feelings and emotions expressed by human beings in their language. Text sentiment analysis is one of the most important parts of sentiment analysis, which can deal with sentimental subjective texts using NLP and text mining techniques. Sentiment analysis tasks can be divided into the chapter level, sentence level, and word or phrase level according to the granularity of analysis. This study was conducted from the word or phrase level.

The first step of this study was to obtain the Twitter data relevant to the people's opinions on the AVs. As an efficiency scraper for social networking services, "snscrape" (https://github.com/JustAnotherArchivist/snscrape) was used to extract valid data based on several AV related keywords: "autonomous vehicle", "autonomous driving", "autonomous car", "driverless", and "self driving". Each keyword was associated with a large number of search results, and the resulting corpus contained a total of 954,151 tweets from January $1^{st}$, 2019 to November $30^{th}$, 2020 for further pre-processing. This period witnessed the global outbreak of COVID-19 and the further development of AV technologies, which contributed to a large number of relevant tweets during this period. Hence, this period was chosen for investigation in this paper.

The following detailed steps were carried out on the corpus:
(1) Text cleaning: The text part of the tweets contained irrelevant information such as numbers, URLs, special symbols, mentions, emojis, HTML tags, etc., which were filtered out. For hashtags, only the "#" symbol was deleted, and the text content was retained for further analysis. In addition, empty and NaN entries were removed from the dataset.
(2) Stop words removal: Stop words included irrelevant filler words such as *for*, *at*, *on*, *which*, *is*, etc., which were removed from the corpus since these words could not accurately reflect the user's comment attitudes.
(3) Removal of retweets: Re-tweets were the tweets' replies. These replies were supposed to be filtered out because most of them were consistent in attitude with the existing tweet.

The next step was to perform the sentiment analysis to identify emotional attitude (i.e., positive or negative) and the subjective degree (i.e., subjective or objective) in each given tweet text. Two text processing tools of Python, TextBlob (https://textblob.readthedocs.io/en/dev/) and IBM Watson (https://ibm.com/watson),

were adopted to compute the sentiment and subjectivity for the preprocessed text, respectively. Sentiment scores ranged from -1 (most negative) to 1 (most positive), among which 0 indicates a neutral attitude. Subjectivity scores ranged from 0 (most objective) to 1 (most subjective) and were further classified into four categories, including "very objective", "objective", "subjective", and "very subjective", as illustrated in Table 1.

Table 1. Classification and distribution of the subjective score

| category | very objective | objective | subjective | very subjective |
|---|---|---|---|---|
| subjective score | 0~0.25 | 0.25~0.5 | 0.5~0.75 | 0.75~1 |
| proportion | 27.9% | 31.7% | 33.9% | 7.5% |

Then, 53 potential impacting factors that may affect the public's attitudes towards AVs were sorted out, deriving from seven categories: "event", "people", "vehicles", "roads", "environment", "autonomous driving-related companies", and "autonomous driving-related characteristics". Each category included a number of subdivided potential influencing factors, and then each factor needed to have detailed words for further interpretation, which are listed in Table 2. These seven categories were selected considering the role of AVs in the seven elements of transportation.

For Topic 1 (events), the public's opinions relevant to the measures imposed by the governments and COVID-19 were collected. With reference to Topic 2 (people), due care was taken to mainly discuss the human attributes and driver behavior. Detailed words such as "brake", "accelerate" and "drunk driving" constituted the major part of the driver's status and behavior. As for Topic 3 (vehicles), detailed words such as "truck", "speed", and "lidar" showed attention which was drawn to obtaining comments relative to the vehicle types, driving characteristics, and vehicle equipment. As for Topic 4 (roads), different road types and road nodes were considered to explore the public views on the applicability of AVs in different road conditions. Detailed words such as "weather", "morning", "afternoon", "traffic signal" etc., were added in Topic 5 (environment) to discuss the potential relationships between environment-related factors and AVs. Topic 6 mainly covered the AVs-related companies including "Tesla", "Baidu", "Waymo" etc., which may largely reflect the public's attitude towards AV. Topic 7 (AV-related characteristics) focused on capturing viewpoints relative to the possible problems caused by AV, such as safety concerns, legal issues, privacy issues, mobility, congestion, etc.

The detailed words were used to mine the massive tweets as the basic data of the subsequent mathematical-statistical analysis model. If a piece of comment data contained the corresponding word, it was marked as 1, while if it did not contain the corresponding word, it was marked as 0. Thus, every independent variable in the model was a binary variable.

Table 2. AV-related topics, variables, and detailed words from Twitter data

| Topics | Variables | Detailed words |
|---|---|---|
| events | policy | policy; policy publication; |
| | testing | testing; test scenario |
| | COVID-19 | coronavirus; COVID-19; COVID19; epidemic; lockdown; quarantine |
| people | pedestrian | pedestrian; passerby |
| | stress | stress; easy; relaxed; convenience; nervous; tension; anxious |
| | passenger | passenger; chauffeur |
| | car following | car following; follow the car |
| | brake | brake; slow down |
| | accelerate | accelerate; speed up; acceleration |
| | drunk driving | drunk; zonked; stoned; drink-driving; intoxicated |
| | fatigue driving | fatigue driving; tired |
| | sleepy | sleep; sleepy; drowsy |
| | male | male driver |
| | female | female driver; woman driver; chauffeuse |
| | young | young man; stripling; teenager |
| | old | old; elderly |
| | income | income; earning; salary; afford; price; expensive |
| vehicles | truck | truck; wagon; van; lorry |
| | bike | bike; bicycle |
| | speed | speed; velocity |
| | lidar | lidar; radar |
| | blind spot | blind spot; vision blind area |
| roads | highway | highway; expressway |
| | roadway | roadway |
| | urban road | urban road |
| | roundabout | roundabout |
| | toll station | toll station; toll gate; services station |
| | ramp | ramp |
| | intersection | intersection |
| environment | weather | weather; sunny; rainy |
| | morning | morning |
| | noon | noon |
| | afternoon | afternoon |
| | evening | evening |
| | traffic signal | traffic light; signal |
| AV-related companies | Baidu | Baidu |
| | Uber | Uber |
| | Volvo | Volvo |
| | Zoox | Zoox |
| | Voyage | Voyage |
| | Waymo | Waymo |
| | Argo AI | Argo AI |
| | Tesla | Tesla |
| AV-related characteristics | mobility | mobility; convenient; |
| | parking | parking |
| | energy | energy conservation; emission reductions; environmentally friendly; emission fuel efficiency; fuel economy; energy efficiency |
| | congestion | congestion |
| | safety | safe; risk; security; crash; accident; collision |
| | legal issues | legal issues; legal liability; liability issues |
| | privacy | privacy; personal data |
| | public transportation | bus; public trans; metro; subway |
| | cyber issues | cyber security; network security; internet security; hack |
| | ethical issues | ethical; moral |

*Note: Detailed words searches are not case sensitive.*

**2.2 Random forests**

In this study, the Random forests (RF) algorithm was used to choose variables by calculating their importance. RF is an ensemble learning method for classification or regression by constructing a multitude of decision trees (Breiman, 2001). One of the advantages of RF is the power of handling large data sets with hundreds of input variables and mitigate the multicollinearity problem (Yu et al., 2019a). In addition, RF can output the importance of variables (Xu, 2021). Thus, it had been regarded as one of the handiest dimensionality reduction methods for variable selection.

Since the dependent variable in this study (i.e., sentiment scores) was a continuous variable, so the decision trees constructed in RF were regression trees. Three methods are combined in the RF algorithm, including bootstrapping, boosting, and bagging (Yu et al., 2019b). Bootstrapping refers to random sampling with replacement. Bagging is a method that can calculate multiple models at the same time, which can realize parallel calculation and improve the robustness of the model. Boosting can help to reduce bias and avoiding overfitting.

The basic process of Random forests consists of the following three steps (Li et al., 2020c). Firstly, $n_{tree}$ sets of bootstrap databases are formed according to the original database, where $n_{tree}$ is the number of decision trees in the forest. Each set of bootstrap databases randomly selects around two-thirds of samples from the original database with replacement, and bootstrap databases have the same dimension as the original one. Those leaf-out samples in each bootstrap database are called "out-of-bag" (OOB) data. Secondly, each set of bootstrap databases is used to grow an unpruned regression tree, and each split within each tree is created by trying $m_{try}$ candidate variables, where $m_{try}$ denotes the number of different independent variables tested at each node. Lastly, the accuracy of each tree will be calculated by OOB data for each set of bootstrap databases, and the whole accuracy of RF is the average accuracy of all the trees.

During the modeling process, RF can calculate the importance of variables by the mean decrease in Gini (i.e., Gini importance), which is a crucial feature for further selecting variables (Li et al., 2020a). Gini importance measures how much a variable can decrease the impurity of the final model, which is calculated by the weighted average value of the impurity reduction (i.e., variance reduction for regression) at all the splits with this variable across all the trees, as shown in the following equation (Louppe et al., 2013):

$$IMP_{Gini}(x_l) = \frac{1}{n_{tree}} \sum_{n=1}^{n_{tree}} \sum_{t \in n: v(s_t) = x_l} p_t \Delta i(s_t, t) \tag{1}$$

Where $IMP_{Gini}(x_l)$ denotes the Gini importance of the independent variable $x_l$; $p_t$ is the weight, namely, the proportion of samples reaching the $t$th node; $\Delta i(s_t, t)$ is the impurity decrease at the split $s_t$ of the $t$th node; $n_{tree}$ is the number of trees in the forest; $n: v(s_t) = x_l$ denotes that the variable $x_l$ is used at the split $s_t$ in the $n$th tree.

In this study, the "RandomForestRegressor" function in the "scikit-learn" package of Python (version 3.7) was used to build the RF model. For training the model, 70% of the samples were randomly selected and used, while the remaining 30% samples were used to assess the quality of the model. All the parameters were tuned based on accuracy. When the accuracy reached the stable and maximum values, the parameters were determined: the number of trees was 50, the number of variables tested at each split was 7, and nodes were expanded until being pure or containing less than 2 samples.

**2.3 Linear mixed model**

This study employed the linear mixed model to investigate the impacting factors on the public's attitude towards AVs with considering the unobserved heterogeneities caused by the subjectivity of tweets. Considering the fact that simple linear regression ignores possible correlations between observations in the data, a linear mixed model was used to mitigate this problem (i.e., unobserved heterogeneities). The linear mixed model is a statistical technique that accounts for within- and between-subject variance for repeated measures data by providing both unbiased estimates of fixed effects and unbiased predictions of random effects (Jermakian et al., 2017). This approach is a flexible and widely used tool to calculate maximum likelihood estimations for hierarchical, longitudinal, or correlated data analyses (Wang et al., 2017a; Yu et al., 2019a; Yu et al., 2021).

A linear model was employed to model the continuous outcome (e.g., sentiment scores) using the "lme4" package in the statistical software R (version 4.1.0), where odds ratios of fixed effects were analyzed to show relative likelihoods. The process of the general linear mixed model is calculated as below:
$$y = X_i\alpha + Z_i b_i + e_i \quad (2)$$
Where $y$ donates the sentiment scores of tweets related to AVs, which is the dependent variable; $X_i$ denotes the fixed effects; $\alpha$ denotes the coefficient of the fixed effects; $Z_i$ denotes the subjectivity of tweets related to AVs, which is selected as the random effect; $b_i$ denotes the coefficient of the random effect; $e_i$ denotes the observation level error terms.

Firstly, the top 20 variables in the RF regression algorithm with variable importance were used to be the fixed effects of the linear mixed model. Then, according to the results of the first model run, each variable was tested for statistical significance at the $p$ 0.05 level. All the less significant variables ($p>0.05$) were filtered out, and the remaining variables were selected as the input of the next model. Ultimately, the final model was checked on the goodness of fit as suggested by the Akaike Information Criterion (AIC) and the Bayesian Information Criterion (BIC). AIC represents the relative amount of information loss caused by a given model, and BIC estimates a posterior probability function of a model being true (Burnham and Anderson, 2004). The AIC and BIC can be calculated as below:
$$AIC = 2k - 2\ln \hat{L} \quad (3)$$
$$BIC = k \ln(n) - 2\ln \hat{L} \quad (4)$$
Where $k$ donates the number of estimated parameters in the model; $\hat{L}$ donates the maximum value of the likelihood function for the model. $n$ donates the number of observations.

## 3. Results
### 3.1 The distribution of attitudes towards AVs

The average of all tweets' sentiment scores in each month was plotted in Figure 1 that shows the distribution of the public's general attitudes and acceptance towards AVs during the selected period (from January 2019 to November 2020). The maximum value of the average sentiment scores was 0.138 (January 2020), and the minimum value was 0.096 (March 2021). As time changes, the overall trend of the public's attitudes on AVs was relatively stable. The most dramatic fluctuation occurred in March 2021, which was a 24.6% decrease compared to February 2021, which might be affected by the outbreak of the COVID-19. Overall, the distribution of average sentiment scores reflected that the public was slightly optimistic about the prospects of AV technologies. In addition, the number of tweets related to AVs each month was more than 30,000, which showed that AV technologies maintain a hot topic of public discussion. Such large sample data also contributed to improving the effectiveness of our subsequent analysis results.

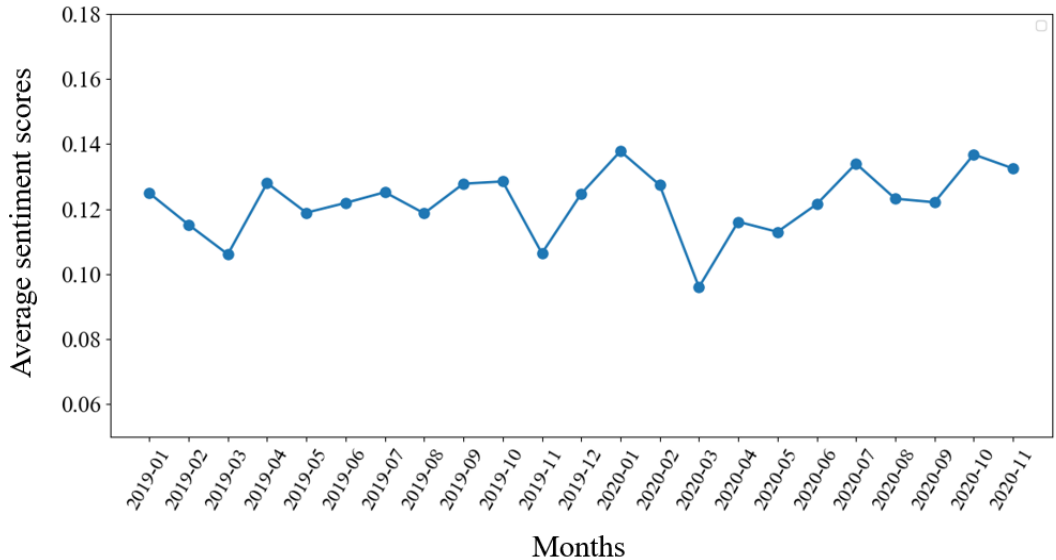

**Figure 1. Distribution of public attitudes towards AVs**

**3.2 Variable selection by Random forests**

The RF algorithm was employed to select the most important candidate independent variables for further analyses. The dependent variable of the RF model was the sentiment scores of tweets related to AVs, which was a continuous variable. The candidate independent variables were 53 potential impacting factors which were listed in Table 2. In this RF model, the importance of all these variables that may affect the public's attitudes towards AV was calculated and ranked.

After calculating the variable importance, the word-cloud function in "wordcloud2" (a visualization toolkit of python) was used to generate the word cloud representation in Figure 2. This visual representation provided a view of all the candidate impacting factors, in which the importance of each variable was defined by its size (the larger, the more significant). The importance of some variables was relatively prominent, such as "mobility", "blind spot", and "Tesla", etc. The correlation between these variables and sentiment scores would be further analyzed through a linear mixed model.

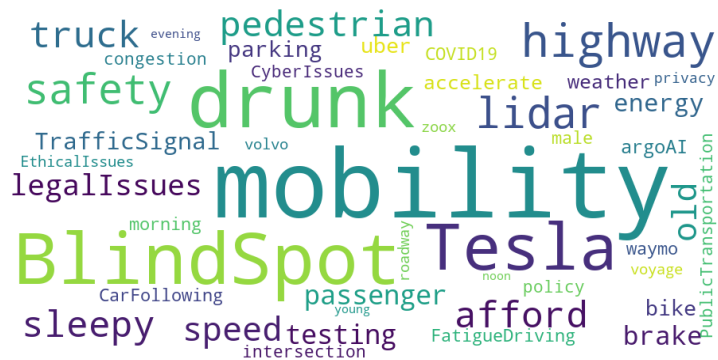

**Figure 2. Word cloud of potential impacting factors**

In addition, Figure 3 illustrates the top 20 variables ordered by importance. It is obvious that "mobility" was the top-ranked one with the variable importance of 0.202.

The advent of AV technologies can further enhance public mobility (Zhao and Malikopoulos, 2019), which greatly affected people's attitudes towards it. Then, "drunk driving", "blind spot", and "Tesla" were ranked second to fourth with the variable importance of 0.121, 0.105, and 0.084. The importance of the remaining 16 variables ranged from 0.011 to 0.037.

From another perspective, 8 of these 20 variables belonged to Topic 2 (people), indicating that human attributes and driver behavior were the hot spots of public attention to AV. Topic 7 (AV-related characteristics) contained many important factors including "mobility", "safety", "legal issues", and "energy", indicating that while AV technologies bring benefits to people, it also brings some legal and environmental conflicts. In addition, "lidar", "truck", "blind spot", and "speed" were three important factors that belonged to Topic 3 (vehicles). Variables such as "COVID-19", "testing", "highway", "traffic signal", and "Tesla" also showed strong importance.

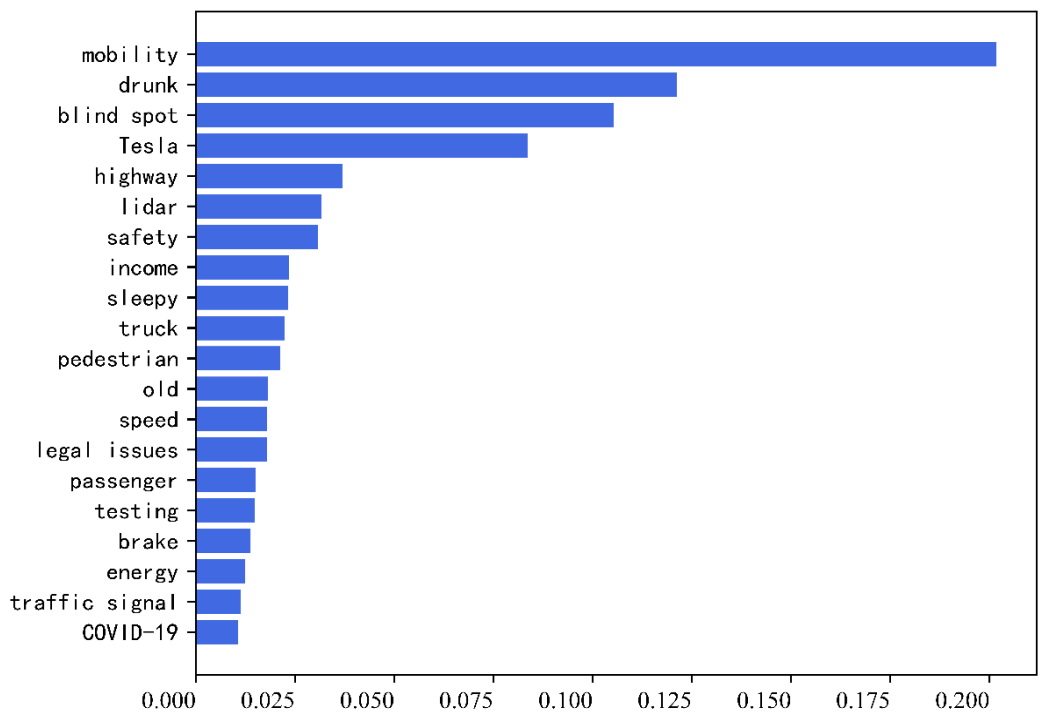

**Figure 3. Variable importance in Random forests (top 20)**

## 3.3 Impacting factors on the public's attitude towards AVs using a linear mixed model

Figure 4 shows the distribution of the sentiment scores grouped by the subjectivity of tweets. The boxplot displays variation in samples of a statistical population, and the results showed that there were a large number of outliers in the "very objective" group without obvious skewness. In the "very objective" group, 86.1% (310,026 in 360,148) of the sentiment scores were 0, representing neutral attitudes. The distribution of the "very subjective" group had fewer outliers, and the median was about 0.2. In addition, the medians of the "objective" group and the "subjective" group were 0.14 and 0.35 respectively. These three groups showed a slightly positive attitude towards AVs.

Additionally, Kruskal Wallis tests by ranks (the nonparametric equivalent of one-way Analysis of variance (ANOVA)) were employed to examine whether there were significant differences among the distribution of sentiment scores in these four subjectivity groups (Feng et al., 2018). The results showed that the distributions of the sentiment scores in four groups were significantly inconsistent ($p<0.01$). Thus, it is necessary to treat subjectivity as a random effect in a linear mixed model.

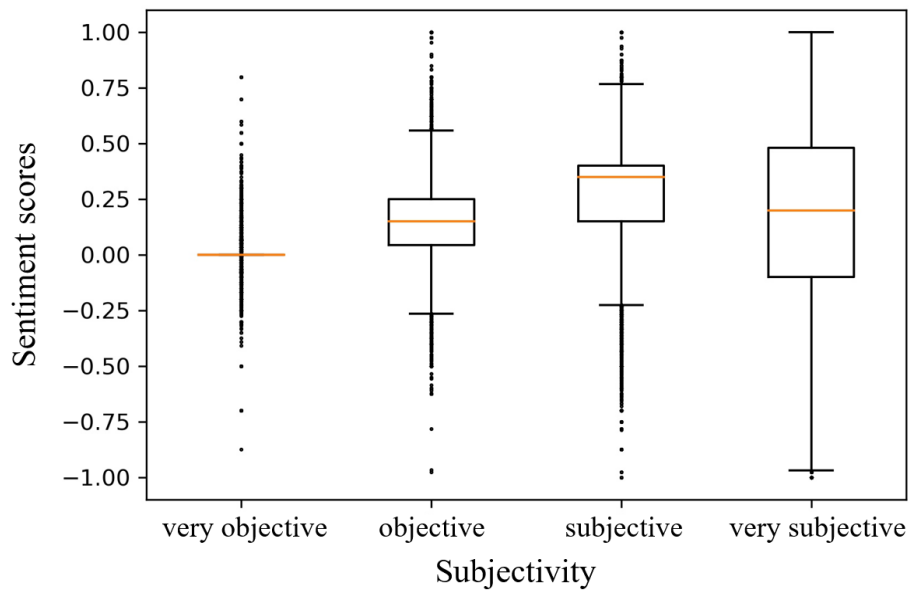

**Figure 4. Distribution of the sentiment scores grouped by the subjectivity of tweets**

To further investigate the significant impacting factors on sentiment scores, a pre-experiment was carried out, in which the top 20 variables ranked in descending order of variable importance were used as the inputs of the linear mixed model. After eliminating the insignificant variables ($p>0.05$), in the final linear mixed model, ten variables were chosen as fixed effects, including an AV-related company ("Tesla"), an AV-related characteristic ("mobility"), a road-related variable ("highway"), people-related variables ("drunk", "sleepy", and "pedestrian"), an event-related variable ("COVID-19"), and vehicles-related variables ("lidar", "blind spot", and "speed"), while the subjectivity of tweets was selected as a random effect.

Figure 5 shows the distribution of the sentiment scores of tweets containing each fixed effect by a violin plot. The violin plot is a method of plotting numeric data, which is more informative than a plain box plot by showing the full distribution of the data with the addition of a kernel density plot on each side (*NITS*, 2021). In this figure, *N* is the number of tweets with each fixed effect. Among the 18,667 tweets relevant to "mobility", the sentiment scores were mostly larger than 0, which had the maximum median value of 0.27. Negative attitudes were expressed more frequently than the other emotions towards the variable "drunk" (*N*=2,723). In terms of the "blind spot" (*N*=659), the majority of Twitter users were skeptical and worried about it. "Tesla" gained most people's attention, due to the number of tweets mentioning Tesla was the largest, 79,678 in total. The median value of sentiment scores of the tweets related to "highway" (*N*=4,842), "lidar" (*N*=17,126), and "speed" (*N*=8,536) were all larger than 0 (0.07, 0.15, and 0.1, respectively). As for the tweets related to "sleepy" (*N*=3,973), "pedestrian" (*N*=6,882), and "COVID-19" (*N*=7,009), the median value of sentiment scores of these tweets were all 0.

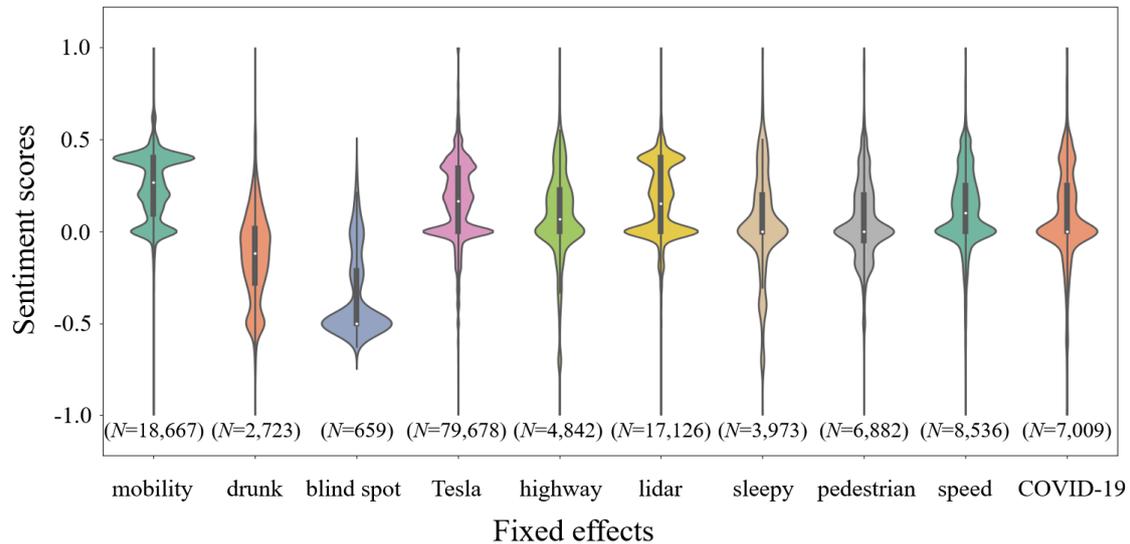

**Figure 5. Distribution of sentiment scores of fixed effects related to AVs from Twitter data**

The results of the final linear mixed model are illustrated in Table 3. The "mobility" was significantly positively associated with sentiment score (t=48.12, *p*<0.01). Most people hold the same view that AV technologies contributed to improving daily commuting ability, especially for people with disabilities (Bennett et al., 2019). Drunk (t=-79.42, *p*<0.01) and blind spot (t=-72.29, *p*<0.01) were significantly negatively associated with sentiment scores. It indicated that the public had a low tolerance for drunk driving, which was illegal even if on AVs occasion. As for the variable "sleepy" (t=-23.84, *p*<0.01), it was also negatively correlated with the dependent variable. People were worried that the AVs system would lead to drivers' excessive trust so that they completely abandoned the control of the vehicle. There were some tweets expressing strong dissatisfaction about the situation where the driver fell asleep when the autopilot function was activated on the highway.

In addition, due to the great progress in the technology of vehicle-mounted sensor equipment such as millimeter-wave radar and high-definition cameras, the variable "lidar" (t=19.40, *p*<0.01) was positively correlated with sentiment scores. However, the "blind spot" remained a public concern (t=-72.29, *p*<0.01). In most AVs solutions, due to the limitations of the lidar vertical field of view range and the overhead installation method, there will be a perception blind spot that is difficult to cover by lidar in the near field area around the body. Potential low obstacles such as pets and children could be extremely risky.

As the vulnerable group in the road environment, "pedestrians" (t=-25.55, *p*<0.01) generally had low trust in AV from the perspective of their safety. Although vehicles were equipped with increasingly sophisticated safety and crash avoidance technology, pedestrian fatalities have risen slightly (Combs et al., 2019). As for the AV-related companies, relevant tweets further demonstrated that "Tesla" (t=10.36, $p<0.01$) contributed to increase the public's favorability and trust in AVs. Further, "highway" (t=-15.10, $p<0.01$) was negatively associated with sentiment scores. Despite extraordinary efforts from many pioneers in tech and automaking, fully AV technologies on the highway were still out of reach except in special trial programs. In terms of speed (t=-5.46, $p<0.01$), the public was worried about the high speed advertised by AVs.

Restrictions on public activities have been in place because of "COVID-19" (t = -13.73, $p<0.01$), resulting in significant changes in mobility patterns and the stagnation of autonomous vehicles tests. It could be the reason making people generating negative attitudes towards AVs during the pandemic period.

The public also used Twitter to report accidents or casualties caused by AV, due to which artificial intelligence and AV had always faced tough questioning. This study set variable "safety" which included detailed words such as "crash", "accident", "injured", etc. However, the influence of "safety" on the public's attitudes towards AVs was not significant ($p>0.05$), so it was not contained in the final model. The public may have mixed feelings about the safety benefits of AVs. On the one hand, AV technologies have the potential in improving traffic safety. On the other hand, the public still worries about the safety of current AV technologies.

Table 3. Linear mixed model results

| Fixed effects estimates | | | | | |
|---|---|---|---|---|---|
| Fixed effects | Estimate | Standard Error | DF | t-value | Pr > \|t\| |
| mobility | 7.294e-02 | 1.516e-03 | 9.541e+05 | 48.12 | < 0.01 |
| drunk | -3.123e-01 | 3.932e-03 | 9.541e+05 | -79.42 | < 0.01 |
| blind spot | -5.759e-01 | 7.966e-03 | 9.541e+05 | -72.29 | < 0.01 |
| Tesla | 7.884e-03 | 7.608e-04 | 9.541e+05 | 10.36 | < 0.01 |
| highway | -4.450e-02 | 2.948e-03 | 9.541e+05 | -15.10 | < 0.01 |
| lidar | 3.057e-02 | 1.576e-03 | 9.541e+05 | 19.40 | < 0.01 |
| sleepy | -7.760e-02 | 3.256e-03 | 9.541e+05 | -23.84 | < 0.01 |
| pedestrian | -6.316e-02 | 2.472e-03 | 9.541e+05 | -25.55 | < 0.01 |
| speed | -1.215e-02 | 2.224e-03 | 9.541e+05 | -5.46 | < 0.01 |
| COVID-19 | -3.365e-02 | 2.451e-03 | 9.541e+05 | -13.73 | < 0.01 |
| **Random effect estimates** | | | | | |
| Random effect | | Variance | | Standard Deviation | |
| Subjectivity | | 0.009897 | | 0.09948 | |
| Residual | | 0.041745 | | 0.20431 | |

In addition, a linear regression model (without considering random effects) was employed by using the "lm" function in the statistical software R (version 4.1.0) to be compared with the linear mixed model. As demonstrated in Table 4, the AIC and BIC of the linear mixed model were -322,623 and -322,459 which were much lower than that in the linear regression model (-124,912 and -124,759 for AIC and BIC, respectively). It indicated the superiority of the linear mixed model.

**Table 4. Comparison of the linear mixed model and linear regression model**

|  | AIC | BIC |
|---|---|---|
| Linear mixed model | -124,912 | -124,759 |
| Linear regression model | -322,623 | -322,459 |

## 4. Discussion and Conclusion

This study aims to propose a method by using large-scale social media data to investigate key factors that affect the public's attitudes and acceptance towards AVs. A total of 945,151 Twitter data related to AVs and 53 candidate variables from seven categories were extracted using the web scraping method. Then, sentiment analysis was used to measure the public's attitudes on AVs by calculating sentiment scores. Random forests algorithm was employed to preliminarily select candidate independent variables according to their importance, while a linear mixed model was performed to explore the significant impacting factors on public attitudes towards AVs considering the unobserved heterogeneities caused by the subjectivity level of tweets.

In this study, the use of social media data provided an opportunity for the collection of comprehensive information that might be representative enough for the public's attitudes towards AVs. Traditional survey methods such as field investigation and online questionnaires were limited in understanding the opinions of the mass due to the lack of comprehensive and effective information retrieval and collection. However, extracting data from social media platforms can be considered as an alternative method, for its large quantity, timeliness, and effectiveness. It was found that 56% of individuals ranging in age from 18 to 56 years old preferred to express thoughts and opinions on social media such as posting and commenting on updates, rather than questionnaires (Whiting and Williams, 2013).

Among all the significant impacting factors explored by the linear mixed model on the public's attitudes towards AVs, "mobility" was the most important one according to the results of RF. A study found that people's awareness of mobility-related developments can increase the acceptance of driverless shuttles (Nordhoff et al., 2018). Increased mobility especially for vulnerable people was one of the main perceived advantages of AVs. The elderly and the disabled will be able to better obtain medical services and participate in society with improved mobility (Yang and Coughlin, 2014). Moreover, the higher mobility brought about by AVs in the future will enhance people's ability to travel from one place to another, which may lead to people's willingness to live farther away from the city. Faced with these changes, it may cause a chain effect and be a challenge for the population distribution, education structure, and transportation system, and so on. It is worth thinking about by policymakers.

This study also found that the public was more inclined to express a negative attitude towards AVs when the public expressed themselves with the following words:

"drunk", "sleepy", "pedestrian", "speed", and "blind spot". Thousands of fatalities were attributable to drunk driving every year despite the attempts made to warn and educate drivers by the government in the methods of propaganda and even criminal punishment. Although negative attitudes were expressed by the public towards drunk driving in AVs occasion, it remained a question whether we should punish someone who is under alcohol's influence riding in the driver's seat of an autonomous car in the same way that we would punish someone who is controlling a vehicle (Hanna, 2014).

Drivers were also strictly forbidden to sleep while driving in the current legal system. It can be found in this research that the public was opposed to this behavior when AV technologies had not been fully popularized. However, drunk or sleepy drivers may be allowed to drive autonomous vehicles when full driving automation is realized. In addition, the efforts of the legislatures are needed to decrease the public's concerns. The redefining of the word "operate" is supposed to be considered by the state legislatures (Douma and Palodichuk, 2012). Traditional laws believe that the operator of a motor vehicle actively controls the vehicle, which is not the case with AVs.

The public consideration of pedestrian safety issues and the speed of AVs have also affected their attitudes towards AVs. The pedestrian fatalities have risen slightly (Combs et al., 2019) despite the high technologies and new theories showing up (Mahadevan et al., 2018), which have been widely applied on the AVs. Policymakers are supposed to take more considerations related to pedestrian safety when formulating traffic policies regarding AVs. There are still many people who refused to sit in high-speed vehicles controlled by software. However, AVs contributed to the reduction in speed variance in mixed traffic conditions with both AVs and human-driven vehicles, which help to decrease the probability of collisions (Khondaker and Kattan, 2015). Therefore, people's understanding of AVs needs to be guided, since there is an amount of uncertainty before the actual integration of full AV technology occurs.

The variable "lidar" was found to be positively correlated with sentiment scores in this study, while the "blind spot" showed a negative relationship with the public's attitudes towards AVs, which seemed to be a contradiction. Vehicle equipment such as 3D lidar, stereo vision cameras, and thermal cameras can be helpful in pedestrian recognition and tracking (Wang et al., 2017b; Chen and Huang, 2019). However, the hidden danger of blind spots has not been eliminated yet. The limitations of the lidar and relevant equipment can lead to blind spots such as low obstacles (e.g., pets and children). Lidar manufacturers should increase investment in the research and development of lidar products, aiming for better performance, smaller size, and lower cost, which performed as an important guarantee for the safety of AVs.

Tesla is one of the pioneers to push the AVs into the market and has recognized the need to shift its innovation from the mechanical parts of the car to its electronics and software (Mallozzi et al., 2019). Positive attitudes were found in 65% of tweets mentioning Tesla in this study due to its contribution to the development of AVs. However, sufficient supervision is still needed for electric vehicles companies like Tesla, since they have been an important market segment. Private companies always tend to exaggerate the benefits of their products, the autopilot functionality of Tesla is claimed to meet the full self-driving capabilities in the future with software updates designed to improve functionality over time. An objective understanding of AVs is needed for consumers that current AV products still need further testing by the market.

In terms of the factor "COVID-19", it has created a spreading and ever-higher healthy threat to people and the manufacturing system which incurred severe disruptions and complex issues to industrial networks (Li et al., 2020b), thus making people generating negative attitudes towards AVs during the pandemic period. However, this period witnessed the development of autonomous delivery vehicles that had the potential to radically change the way groceries are delivered to customer homes (Kapser et al., 2021). A higher acceptance towards AVs can be predicted as more needs in the life of people are meet by AVs.

The results of this study are beneficial to policymakers, automotive industries, technology companies, and general consumers. It can help policymakers and legislators create plans and designate new laws based on public opinions. Developers and manufacturers of AVs can rethink their commercial strategy and product positioning according to the voice of customers. In addition, this research also proposed a complete and efficient technical route from social media platform data collection, data processing to modeling analysis at a low cost, which can be applied to other transportation-related themes. In future work, other information, such as personal information (age, education level, gender, etc.), locations, and so on, and social media data from other platforms will be included and compared to improve the effectiveness of the model.

**CRediT authorship contribution statement**
**Shengzhao Wang:** Conceptualization, Data curation, Methodology, Writing - original draft. **Meitang Li:** Data curation, Methodology, Software, Validation. **Bo Yu**: Conceptualization, Data curation, Methodology, Writing - original draft. **Shan Bao:** Conceptualization, Data curation, Methodology, Writing – original. **Yuren Chen:** Conceptualization, Funding acquisition, Writing - original draft.

**Declaration of competing interest**
The authors declare that they have no known competing financial interests or personal relationships that could have appeared to influence the work reported in this paper.

**Acknowledgment**
This project was supported by the National Key Research and Development Program of China (No. 2017YFC0803902) and Shanghai Intelligent Science and Technology Category IV Peak Discipline.